\documentclass[12pt]{article} 
\usepackage[margin=1in]{geometry}
\usepackage{hyperref}
\usepackage{amsmath}
\usepackage{setspace}
\usepackage{slashed}
\usepackage{amssymb}
\usepackage{color}
\usepackage{url}
\usepackage{epstopdf}
\usepackage{authblk}
\usepackage{siunitx}
\usepackage{graphicx}
\usepackage{wrapfig}
\usepackage[labelfont=bf]{caption}

\usepackage[authoryear,round]{natbib}

\title{The Fate of Causal Structure under Time Reversal}
\author{Porter Williams\\
	Department of Philosophy\\
	University of Southern California}
\date{}                     %% if you don't need date to appear
\begin{document}
	\maketitle

\begin{abstract}
\noindent What happens to the causal structure of a world when time is reversed? At first glance it seems there are two possible answers: the causal relations are reversed, or they are not. I argue that neither of these answers is correct: we should either deny that time-reversed worlds have causal relations at all, or deny that causal concepts developed in the actual world are reliable guides to the causal structure of time-reversed worlds. The first option is motivated by the instability under intervention of time-reversed dynamical evolutions. The second option is motivated by a recognition of how contingent structural features of the actual world shape, and license the application of, our causal concepts and reasoning strategies.
\end{abstract}

\section{Introduction}

Bertrand Russell famously declared that causal notions in fundamental physics were a ``relic of a bygone age'' \citep{1912_russell}. His most influential argument for this conclusion invoked the time-symmetric character of dynamical evolution in fundamental physics and has come to be called the \textit{Directionality Argument}. The argument, in short, begins with the claim that the relationship between cause and effect is asymmetric in ways that cannot be grounded in time-symmetric dynamical laws. Since the dynamical laws of fundamental physics \textit{are} time-symmetric, they cannot satisfactorily distinguish cause from effect: if $A \, \longrightarrow \, B$ in one temporal direction, then $B \, \longrightarrow \, A$ in the other temporal direction. Russell concludes that fundamental physical theories cannot ground causal relations. 

Russell is far from alone in believing that reversing temporal order also reverses causal order. Of course, anyone who concludes with Russell that the Directionality Argument shows that causal relations cannot be grounded in fundamental physical theories certainly believes it. Even those who explicitly reject the conclusion of the Directionality Argument believe it on other grounds. For example, any advocate of transference theories, like \citep{1984_salmon,1994_salmon} or \citep{2000_dowe}, will certainly believe that reversing time reverses causal relations: a causal interaction between A and B \textit{just is} the transfer of a conserved quantity, like energy-momentum, from A to B. The time-reverse of that dynamical evolution, of course, will involve the transfer of energy-momentum from B to A.

In a similar vein, \citep{2009_ney} explicitly endorses an account of ``physical causation'' that is just nomological determination -- a temporally symmetric notion -- and embraces the fact that 

\begin{quote}
$\ldots$ if the universe does not have any fundamental, built-in temporal asymmetries, this seems to be what we are left with. There is still causation, because there is still physical determination. But the distinction between what is the cause and what is the effect may not be fundamental \citep[pp. 752-3]{2009_ney}.
\end{quote}

Meanwhile \citep{2007_price} endorses a quite different account of causation -- a perspectival account according to which causal asymmetries are determined by our psychological perspective as deliberators -- but motivates it by stating that if there were a region of the universe in which entropy was decreasing, ``intelligent creatures would have a time-sense reversed relative to ours$\ldots$ For them, the causal arrow runs directly counter to the way it runs for us'' \citep[p. 273]{2007_price}. Indeed, Price thinks that his belief that causal order will be reversed in time-reversed worlds, combined with the time-symmetric character of the dynamical laws of fundamental physics, ``provides something close to a trump card for perspectivalism'' \citep[p. 269]{2007_price}.\footnote{\label{pricetuning}Price makes an important assumption about an anti-entropic universe: that such a universe does not require any more fine-tuning than an entropic one. As he says, ``at least in the absence of any time-symmetry in the underlying physics, a fine-tuning required is the same in either temporal direction'' \citep[p. 273]{2007_price}. I think this assumption is false, as I will discuss in section 2.}

And \citep[section 3.1.2]{1990_tooley} seems to think that \textit{any} account of causation according to which causal relations supervene on non-causal facts and relations is committed to accepting that reversing temporal order will also reverse causal order.

I think this is mistaken: worlds with reversed temporal order should not be thought of as exhibiting reversed causal relations. To this extent I agree with \citep{2020_farr}. Farr points out that if causal relations in one temporal direction satisfy standard statistical desiderata, like the Causal Markov Condition, they will generally not satisfy those criteria in the reversed temporal order. On the basis of these asymmetries of statistical independence, he concludes that the causal structure of a world is invariant under time reversal:

\begin{quote}
The issues of agency and the [Causal Markov Condition] lead to the same judgements about causal direction regardless of what one takes to be the underlying direction of time. This entails that any underlying time-reversal invariance of the microphysical description is beside the point; one may hold that there is a clear causal direction $\ldots$ which is invariant under time reversal \citep[p. 197]{2020_farr}.
\end{quote}

However, here I part ways with Farr as well: I do not think that reversing temporal order should be thought of as leaving causal ordering invariant. Instead, I think that the correct attitude is one of the following:

\begin{enumerate}
	\item There are no causal relations \textit{at all} in time-reversed worlds.
	\item We have no epistemic warrant for judging that our concepts of causation are reliable guides to the nature, and presence, of causal relations in time-reversed worlds. 
\end{enumerate}

The two answers differ primarily in epistemic audacity. If we are audacious enough to apply to time-reversed worlds the concepts and strategies for causal reasoning that we have developed to help us navigate the actual world,  then we should conclude that there are no causal relations at all in such worlds. Alternatively, upon recognizing how thoroughly foreign to us a time-reversed world is, we might decide that our actual-world standards for judging the nature and presence of causal relations are inapplicable. Ultimately, I think this second answer is the more plausible one: epistemic humility demands that we simply withhold judgment about the presence or absence of causal relations in time-reversed worlds, in light of the restricted scope of our frameworks for causal reasoning.

\section{Reversing Time}

As \citep{2013_farr_reutlinger} have pointed out, the phrase ``time-symmetric character'' that appears in the Directionality Argument is ambiguous and can be precisified in two distinct ways:\footnote{A similar distinction was pointed out by \citep{2002_earman}, but was not applied to the Directionality Argument.}

\begin{enumerate}
	\item \textit{Invertibility}: Given a state of the system $S(t)$, the dynamics $\mathcal{D}_\tau$ of the physical theory determine both the state $S(t + \tau)$ and the state $S(t - \tau)$. 
	\item \textit{Time-Reversibility}: Given a sequence of states of the system $S(t_1), \, \ldots, \, S(t_n)$  related by the dynamics $\mathcal{D}_{t_n}$ and a time-reversal operator $\mathcal{R}$, the time-reversed sequence of states $\mathcal{R}S(- t_n), \, \ldots, \, \mathcal{R}S(- t_1)$ is also related by the dynamics $\mathcal{D}_{-t_n}$.
\end{enumerate}

Both \citep{2002_earman} and \citep{2013_farr_reutlinger} see \textit{Invertibility} as the more fundamental sense in which the dynamics of a theory can exhibit a``time-symmetric character'': Earman because any classical or quantum system admitting a Hamiltonian formulation will satisfy \textit{Invertibility} even if it doesn't satisfy \textit{Time-Reversibility}, and Farr and Ruetlinger because one can understand the two dynamically allowed sequences of states secured by \textit{Time-Reversibility} to be occurring in the same temporal direction.\footnote{Farr and Ruetlinger draw on the same point made earlier by \citep[chapter 4.2]{2007_maudlin}.} I won't make much of the distinction:  in my discussion below I will assume that both conditions are always satisfied. As is evident from the statement of \textit{Time-Reversibility}, I will also adopt the standard interpretation of time reversal according to which time reversal includes a (perhaps non-trivial) operation on the state of a system in addition to changing the sign of the time coordinate.\footnote{A referee wanted justification for this. The appropriate understanding of time reversal has received a lot of philosophical attention in recent years \citep{2000_albert,2000_callender,2002_earman,2004_malament,2009_arntzenius_greaves,2017_roberts,2019_allori,2020_farr,2019_donoghue_menezes,2020_donoghue_menezes,2020_callender,2020_struyve}. The initial stimulation for much of this work were the arguments for a non-standard definition of time reversal by \citep{2000_albert} and \citep{2000_callender}. For reasons compactly summarized in \citep{2019_roberts}, I remain partial to the traditional account and will adopt it throughout this paper.\label{trdef}}

I will also assume that the dynamics of the theory are deterministic. This is sufficient to account for all ``fundamental'' dynamical evolutions of classical mechanics and for unitary evolution in quantum theories.
%\footnote{Of course, we require that the time-reversal operator $\mathcal{R}$ preserve transition probabilities: $\braket{\phi}{\psi} = \braket{\mathcal{R}\phi}{\mathcal{R}\psi}$.}
For simplicity I will stick to the context of classical statistical mechanics, but see \citep{2022_williams} for discussion of how some of the themes here apply to quantum states related by unitary dynamical evolution.

Consider a classical description of the behavior of an \textit{N}-particle system whose dynamical evolution is governed by a Hamiltonian $\mathcal{H}$. This system could be anything: a gas in a box, a basset hound, the City of Pasadena, the entirety of the Pacific Ocean, or the observable universe as a whole. In all of those cases, the \textit{microstate} $S(t)$ of the system is determined by specifying the position and momentum of each particle: 

$$S(t) = \left(q_1(t), \, p_1(t), \, \ldots , q_N(t), \, p_N(t)\right)$$

Dynamical evolutions of this system trace out continuous curves connecting different microstates $S(t_1), \, \ldots, \, S(t_n)$ of the system. The space of all possible microstates of this system is called the \textit{phase space} of the system.

We would also like to describe thermodynamic properties of the system. Roughly speaking, one accomplishes this by partitioning the phase space of the system into \textit{sets} of microstates; we call these sets \textit{macrostates}. All of the microstates contained in a particular macrostate describe a system that exhibits the same values for some set of specified thermodynamic properties like temperature, pressure, total magnetization, etc. It is the values of these thermodynamic properties that define the different macrostates. We can then assign an \textit{entropy} to the system: the value of the entropy of the system at any instant is a function of the thermodynamic properties of the system at that instant. More precisely, the entropy assigned to a system at an instant is determined by the volume of phase space occupied by the macrostate in which the microstate $S(t)$ of the system lies at that instant.\footnote{By \textit{entropy} here I mean \textit{Boltzmann entropy}. Nothing in the paper depends on this choice. For an illuminating and thorough discussion of entropies in classical and quantum mechanics, see \citep{2020_goldstein_et_al}.}

It has been known since the time of Boltzmann that the entropy of closed systems is overwhelmingly likely to increase over the course of their dynamical evolution: for essentially any Hamiltonian $\mathcal{H}$, essentially any non-equilibrium microstate $S(t)$ will evolve to a later microstate $S(t + \tau)$ of higher entropy. However, since Boltzmann's time it has also been known that for essentially any Hamiltonian $\mathcal{H}$, essentially any non-equilibrium microstate $S(t)$ will also evolve to an \textit{earlier} microstate $S(t - \tau)$ of higher entropy: as \citep[p. 77]{2000_albert} puts it, ``the overwhelming majority of the trajectories passing through any particular non-maximal-entropy [macrostate] must just then be in the process of \textit{turning around}.'' This is troubling: the claim that essentially any non-equilibrium microstate $S(t)$ represents a local entropy \textit{minimum} along essentially every dynamical trajectory that passes through it conflicts directly with our memories of the past: isolated gases were not warmer in the past than they are in the present, my eggs were not more scrambled, my hair (sadly) not less grey, and so on. 

However, ``essentially any microstate'' does not mean ``every microstate'': included in the set of microstates that make up any macrostate, there will be a subset of microstates $S_{ab}(t)$ that, under the influence of $\mathcal{H}$, evolve to higher entropy states in one temporal direction and to \textit{lower} entropy states in the other temporal direction. This subset of ``abnormal'' microstates states $S_{ab}(t)$ inhabits a fantastically miniscule portion of the phase space volume occupied by the macrostate in question. Furthermore, this miniscule volume is itself scattered in tiny, geometrically non-uniform clusters and filaments throughout the volume of phase space associated with the macrostate in question; see Figure 1. Given their sparseness, it would seem fantastically unlikely that the present state of the actual world is such a state; nevertheless, all evidence indicates that it is. Providing an account of the foundations of statistical mechanics according to which such a history of the world is not fantastically unlikely, but in fact \textit{highly probable}, has generated a host of conceptual difficulties in the foundations of statistical mechanics \citep{1993_sklar,2000_albert,2007_uffink}.

\begin{wrapfigure}{r}{0.4\textwidth} %this figure will be at the right
	\centering
	\includegraphics[width=0.4\textwidth]{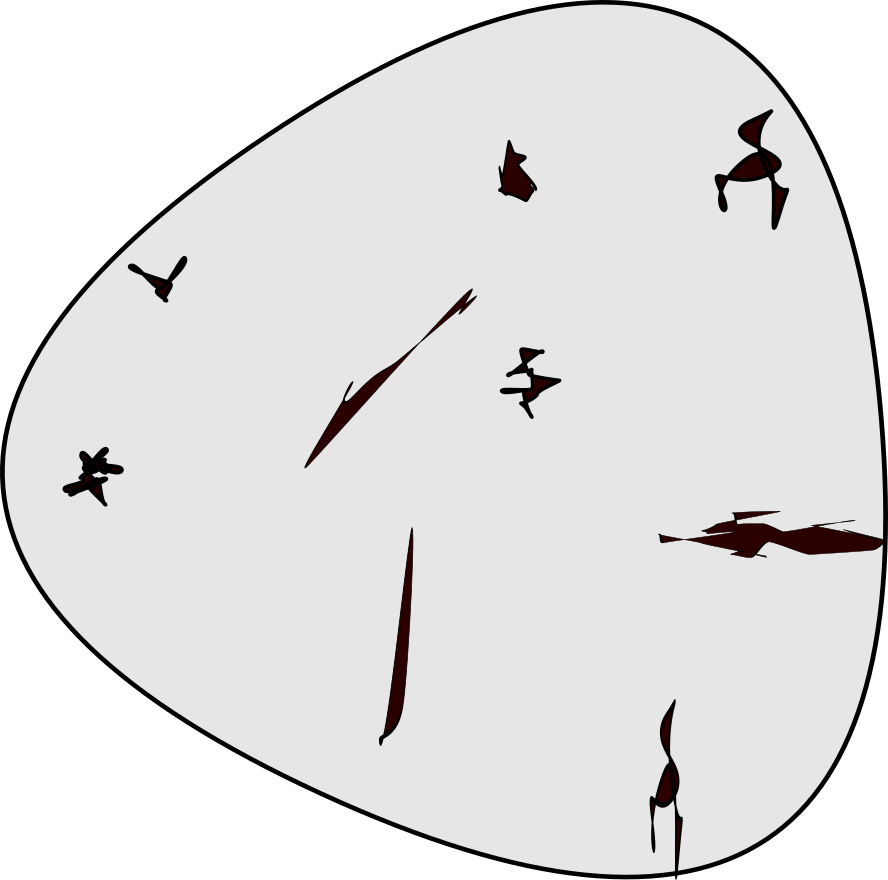}
	\caption{The set of abnormal microstates $S_{ab}$ distributed throughout a macrostate. (Not drawn even close to scale.)}
\end{wrapfigure}

However one secures such an account, the result is that the universe as a whole -- and essentially any closed system within that universe -- dynamically evolves from microstates of lower entropy to microstates of higher entropy. Call worlds like this \textit{entropic}. The actual world is entropic. The time-reverse of an entropic world is a world in which microstates of higher entropy dynamically evolve into microstates of lower entropy. Call worlds like this \textit{anti-entropic}. The time-reverse of the dynamical history of any entropic world corresponds to the dynamical history of an anti-entropic world.

A characteristic feature of entropic worlds is that their identity \textit{as} entropic worlds is very stable. Suppose that any physical system, up to and including the universe as a whole, is in a particular non-equilibrium microstate $S(t)$ that will evolve, under the influence of the Hamiltonian $\mathcal{H}$,  into a microstate $S(t + \tau)$ of higher entropy. As stated above, essentially any \textit{alternative} microstate $S'(t)$ that lies in the same non-equilibrium macrostate as $S(t)$ will also evolve, under the influence of the same Hamiltonian $\mathcal{H}$, into a microstate $S'(t + \tau)$ of higher entropy. No particular coordination, or fine-tuning, of the microstate to the particular Hamiltonian $\mathcal{H}$ is required to secure entropic dynamical evolution. 

It is worth lingering on this point for a moment because elucidating the type of fine-tuning that is, and is not, required to secure entropic dynamical evolution brings out an important difference between entropic and anti-entropic worlds.\footnote{Much of my discussion about the differences in fine-tuning is indebted to \citep[chapter 4.4]{2007_maudlin}.} Many proposals for ensuring that our world lies on a dynamical trajectory  along which low entropy microstates evolve into high entropy microstates posulate something like the Past Hypothesis: that the initial microstate of the universe had very low entropy \citep[chapter 5]{1965_feynman},\citep{2000_albert,2011_wallace}. Macrostates with low entropy occupy a small volume of the phase space of any physical system and the universe itself is no exception, so these proposals require a certain type of fine-tuning: the initial state of the universe had to be ``special'' in the sense that it had to be selected from a small volume of phase space. But that's it: to ensure entropic dynamical evolution, there is no need to choose a \textit{particular} microstate that is fine-tuned to the \textit{particular} Hamiltonian governing the system. In fact, the specific form of the dynamics isn't all that important: the set of microstates that occupy the low-entropy macrostate for the Hamiltonian $\mathcal{H}$ will be essentially the same for a reasonably large neighborhood of different Hamiltonians $\mathcal{H}'$.

In anti-entropic worlds, the situation is importantly different. A characteristic feature of anti-entropic worlds  is that their identity \textit{as} anti-entropic worlds is extremely \textit{fragile}.  The anti-entropic dynamical evolution of a physical system requires a fantastically precise dynamical coordination between the constituent particles of the system: all the particles in my kitchen coordinate to unscramble eggs and launch them back into their shells, the particles in my coffee and mug and local environment cooperate to unmix my coffee and cream, stomach acids conspire to reassemble digested foodstuffs, and so on. If even a few particles fail to play their role, the anti-entropic dynamical evolution falls apart. A few displaced molecules in my frying pan collide with, and disrupt, the carefully coordinated trajectories of other molecules, those molecules fail to displace the appropriate molecules in my scrambled eggs and/or the surrounding air molecules, and so on.\footnote{Similar observations about the fragility of anti-entropic evolution are invoked by \citep{2001_elga} to present a problem for David Lewis's account of the asymmetry of counterfactual dependence. See, in particular, Elga's description of the anti-entropic dynamical evolution of a fried egg.}

We can be slightly more precise about this. If a microstate $S(t)$ lies on a dynamical trajectory along which high entropy microstates evolve into low entropy microstates, then essentially any microstate $S^\epsilon (t)$ obtained as a small perturbation of $S(t)$ will \textit{not} lie along such a dynamical trajectory. That is, if a microstate $S(t)$ is in the set $S_{ab}(t)$ of ``abnormal'' microstates that undergo anti-entropic dynamical evolution under the influence of a particular Hamiltonian $\mathcal{H}$, then the fact that $S_{ab}(t)$ occupies such a fantastically tiny, and scattered, volume within any macrostate means that small perturbations of $S(t)$ will almost certainly \textit{not} be in the set $S_{ab}(t)$. For example, if the microstate 

$$ S(t) = \left( q_1(t), \, p_1(t), \, \ldots , q_N(t), \, p_N(t) \right) $$

lies in the macrostate $\Gamma$ \textit{and} the subset $S_{ab}(t)$ of ``abnormal'' microstates, then the microstate $S^\epsilon(t)$ that results from a small perturbation of the positions and/or momenta of some subset of the particles

$$ S^\epsilon(t) = \left(q_1(t), \, p_1(t), \, \ldots \,  q_i(t) + \epsilon, \, p_i(t) + \epsilon, \ldots, \,  q_m(t) + \epsilon, \, p_m(t) + \epsilon, \ldots, \, q_N(t), \, p_N(t) \right)$$

will still lie in the macrostate $\Gamma$ but will \textit{not} lie in $S_{ab}(t)$. This means that, evolving under the influence of the Hamiltonian $\mathcal{H}$, the perturbed microstate $S^\epsilon(t)$ will generate an \textit{entropic} dynamical evolution. This is one sense in which anti-entropic worlds are fragile: at any instant \textit{t}, small perturbations of the microstate of that world will transform it into an entropic world, i.e. shift it onto a dynamical trajectory along which low entropy microstates evolve into higher entropy microstates.\footnote{This fact is exploited in \citep[chapter 9]{2000_albert} to argue that if a spontaneous collapse theory like GRW is the correct description of quantum behavior, then anti-entropic dynamical evolution is essentially impossible: the spontaneous localizations will take any state initially in the set $S_{ab}(t)$ and very rapidly perturb it into a state that lies outside that set.} I will return to this in section 3.

There is a second type of fine-tuning required in anti-entropic worlds: the microstate generating the anti-entropic evolution has to be delicately tuned to the \textit{particular} Hamiltonian $\mathcal{H}$ that determines that dynamical evolution of the system. Physically, this is because the coordination between all of the particles of a system required for an anti-entropic dynamical evolution depends extremely sensitively on the particular Hamiltonian governing the system. If a microstate $S(t)$ would generate an anti-entropic evolution under the influence of $\mathcal{H}$, then that same microstate will generate an \textit{entropic} evolution under the influence of the Hamiltonian $\mathcal{H}^\epsilon$ obtained from $\mathcal{H}$ by some tiny modifications -- for example, to the range or strength of interactions. Those modifications will (among other things) result in slightly different scattering angles from particle interactions which, in turn, will disrupt the careful coordination of particles secured by $\mathcal{H}$ and progressively wipe out the delicate correlations that are needed between all of the particles at every instant \textit{t} to generate anti-entropic dynamical evolution. Formally, it is because the meaning of ``abnormal'' in the definition of the set of abnormal microstates $S_{ab}(t)$ is \textit{dynamical} abnormality: it is the set of microstates that undergo abnormal dynamical evolution, i.e. evolve from higher entropy microstates to lower entropy microstates. Modifying the Hamiltonian thus completely redefines the set of ``abnormal'' states $S_{ab}(t)$.\footnote{The same point is made by \citep[p. 132-33]{2007_maudlin}.}

The fine-tuning in an anti-entropic world is different not only in degree, but also in kind, from the fine-tuning in entropic worlds associated with proposals like the Past Hypothesis. In entropic worlds one can randomly pick a microstate from a low-entropy macrostate without knowing much at all about the particular Hamiltonian $\mathcal{H}$ determining the system's dynamical evolution, then show that those dynamics will evolve that low-entropy microstate into a high-entropy microstate. The initial microstate is ``special'' in the sense that it comes from a macrostate that occupies a small volume in the system's phase space, but  securing entropic dynamical evolution does not \textit{additionally} require that ``special'' microstate to contain strong correlations between the positions and/or momenta of particles with no antecedent causal connection, nor does picking a ``special'' microstate that will generate entropic dynamical evolution depend sensitively on the details of a particular $\mathcal{H}$. In the terminology of \citep{2020_woodward}, both Causal to Statistical Independence (CSI) and Variable/Relationship Independence (VRI) are satisfied in entropic worlds. In anti-entropic worlds, this is no longer true: the initial microstate describes strong correlations between the positions and/or momenta of particles with no antecedent causal connection, and those correlations must be very precisely tailored to the particular Hamiltonian $\mathcal{H}$ determining the dynamical evolution of the system. In other words, both CSI and VRI fail in anti-entropic worlds.

\section{Causation Reversal?}

What does any of this have to do with time reversal and causal ordering? The answer, in short, is that according to any difference-making account of causation -- any account on which causal claims entail a claim about how an effect would be altered by local alterations of the cause --  there are no causal relations in anti-entropic worlds. This includes interventionst accounts of causation, accounts of causation according to which causal claims reduce to (or non-reductively entail) claims about counterfactual dependence, and so on. If causation is a difference-making relation then inverting the temporal ordering of the states of an entropic world does not result in a world with inverted causal relations: it results in a world with no causal relations at all. 

More precisely, the following two claims are inconsistent:

\begin{enumerate}
	\item At least some causal claims, like \textit{temperature causes pressure} or \textit{smoking causes lung cancer}, are meaningful in world $\omega$.
	\item Closed systems in $\omega$ (including $\omega$ itself) dynamically evolve from states of higher entropy to states of lower entropy.
\end{enumerate}

The argument for the inconsistency is simple, but first recall a couple elementary features of interventionist accounts of causation.\footnote{I will adopt interventionist language from here on out, but the argument would go through just as well with any other difference-making account.} Claims like \textit{temperature causes pressure} mean ``there is some intervention that can be performed on the \textit{temperature} of a physical system that is systematically correlated with a change in the \textit{pressure} of that system.'' And to \textit{intervene} on a variable $\mathcal{V}$ is to bring that variable, and that variable alone, under the total control (or partial control, if one considers soft interventions) of the investigator while holding all other variables fixed. If there is no act one can perform -- even in principle -- to change the value of $\mathcal{V}$ without \textit{that act}, and not the resulting change in $\mathcal{V}$, also changing the value of a distinct variable $\mathcal{Z}$, then one cannot intervene on $\mathcal{V}$. For example, if there was no act that one could perform to change the net magnetization of an iron bar that didn't also change its temperature, or the air temperature in the lab as a whole, or the net magnetization of the iron bar in the lab down the hall, or $\ldots$ then, in such a world, one could not intervene on the net magnetization of an iron bar.\footnote{Of course, in the real world one \textit{can} intervene on the net magnetization of an iron bar precisely by changing its temperature: cool it below its critical temperature. This brings out the difference nicely: one is acting to intervene on the temperature of the bar and it is the change in the temperature that \textit{brings about} a change in its net magnetization. It is \textit{not} whichever specific action one took to lower the temperature that \textit{itself} brought about the change in net magnetization in the iron bar.}

Here, then, is the argument that (1) and (2) are inconsistent. We know from the discussion in section 2 that anti-entropic dynamical trajectories are extremely fragile: if a microstate $S(t)$ generates an anti-entropic dynamical trajectory under the influence of a Hamiltonian $\mathcal{H}$, then states $S^\epsilon(t)$ obtained by tiny perturbations of $S(t)$ almost surely will not. I want to emphasize that by ``tiny perturbations'' I mean perturbations that are completely physically unnatural, let alone attainable: changing by a negligible amount the positions and/or momenta of a few random particles in large system, like a dilute gas, a frying pan, a few cells in a human body, downtown Los Angeles, and so on. Such interventions would require controlling the state of every particle in the system to a physically unattainable degree of precision, but that's not really the issue. Even if they were physically attainable, there is no way to know \textit{which} few particles in the system, if any, one could subject to such a tiny perturbation at a given instant without disrupting the anti-entropic dynamical evolution: that is, without perturbing the system in the state $S(t)$ into a state $S^\epsilon(t)$ that lies outside any of the tiny, geometrically non-uniform regions of phase space that make up $S_{ab}(t)$. And even if one could know, and could perform such a precise intervention, these would not be interventions on anything remotely like meaningful physical properties of a system: there is no physically interesting property corresponding to the small and scattered collection of particles in downtown Los Angeles that one could minimally perturb without disrupting the anti-entropic dynamical evolution of the city as a whole.

Genuine collective physical properties, like a nucleotide sequence in an mRNA molecule, the temperature of an iron bar, or the amount of cigarette tar in one's lungs, are associated with the positions and/or momenta of \textit{enormous} numbers of individual particles. And interventions on those propeties, like mRNA editing, lowering the temperature of an iron bar, or smoking one pack a day rather than two, involve large changes in the positions and momenta of enormous numbers of particles. Those kinds of interventions will inescapably take a state $S(t)$ that lies in $S_{ab}(t)$  and produce a state that lies outside of $S_{ab}(t)$ -- a state that will produce an \textit{entropic} dynamical evolution under the influence of $\mathcal{H}$. Local interventions on physical systems cannot be performed without transforming an anti-entropic world into an entropic one.\footnote{The qualifier ``local'' is important. One could perform interventions on $S(t)$ while maintaining anti-entropic dynamical evolution, but those ``interventions'' would have to be spectacularly non-local. Such an intervention would have to \textit{instantaneously} alter the positions and/or momenta of particles not only in the spatiotemporal vicinity of the intervention, but also in other causally disconnected spacetime regions. The result of a such an anti-entropic-evolution-preserving non-local intervention would be to remove $S(t)$ from one of the ``abnormal'' regions in Figure 1 but to modify it in such a way as to produce a state that lies in one of the \textit{other} ``abnormal'' regions. It is a non-starter to argue that the possibility of ``interventions'' like this would somehow save the day for meaningful causal relations in anti-entropic worlds. For starters, they wouldn't count as interventions in the technical sense at all. (For discussion of a few of the difficulties facing non-local ``interventions'' like this, see \citep{1999_hausmann_woodward}.)}

It may seem incredible that that such small interventions can end up reversing the entropy gradient of the universe, i.e. turning an anti-entropic world into an entropic world. It is clarifying to think about the physical process by which this transformation happens. Suppose that the microstate of the universe is $S(t)$ and that this microstate, under the influence of a Hamiltonian $\mathcal{H}$, will generate anti-entropic dynamical evolution. Among the countless physical states of affairs described by the microstate $S(t)$, my office windows are currently shut. An intervention on the state of the window -- opening it -- will spatially translate the positions of all the particles in the window by some nontrivial amount. The delicate choreography of particle collisions required to secure anti-entropic dynamical evolution will be disrupted: many collisions between air molecules, electromagnetic radiation, dust, molecules in the glass, etc. that needed to take place will not, while many others that would not have occurred now will. The delicate correlations between the positions and momenta of particles in the vicinity of my office window will be  washed out by these collisions, resulting in an \textit{entropic} dynamical evolution. On short time scales, only the collection of particles in a fairly small spatiotemporal region around my office window will undergo entropic evolution; the rest of the universe will continue to evolve anti-entropically.\footnote{In \citep{2001_elga} these spatiotemporal regions of entropic dynamical evolution are called ``infected regions'', although he puts the existence of such regions to somewhat different philosophical ends.} However, at the boundary the particles in this spatiotemporal region will interact with the particles outside of it, disrupting their carefully choreographed dynamical evolutions and thereby expanding the size of the spacetime region in which particles undergo entropic evolution. After enough time, the originally anti-entropic world $\omega$ will have become a world in which every closed system (including the universe itself) undergoes \textit{entropic} dynamical evolution. All because at some earlier time I opened my office window. 

That is the entirety of the argument that (1) and (2) are inconsistent. One cannot perform interventions in anti-entropic worlds without turning them into entropic worlds. If causal claims like \textit{temperature causes pressure} are understood as claims about what would happen to \textit{pressure} under interventions on \textit{temperature} then -- insofar as interventions are incompatible with anti-entropic dynamical evolution -- causal claims cannot be meaningful in anti-entropic worlds. Time-reversing an entropic universe does not produce a universe with reversed causal relations; it produces a universe with no causal relations at all.\footnote{A referee asks whether it makes sense to speak of interventions at all in an anti-entropic world, i.e. whether interventions themselves are necessarily entropy increasing. There are two things to say about this. The first is that an intervention, in the technical sense, is just a change in the value of a variable that has to satisfy certain other conditions (see \citep[chapter 3.1]{2003_woodward} or \citep[chapter 3]{2009_pearl}, for example); it needn't be actually carried out by a human agent. Understood in that way, there are certainly interventions that do not themselves increase entropy: any change to the microstate of a system that does not change its macrostate. How large this class of interventions is will depend on specific details of the physical system and the particular macrostate in question. (Note that this is different than whether the intervention changes the future dynamical evolution of the microstate. As discussed above, it almost certainly will.) The second thing to say is that if interventions were impossible in anti-entropic worlds, then so much the worse for the idea that those worlds contain causal relations.}

\section{Conclusion}

I want to consider two possible avenues of response. The first attempts to save the idea that there are causal relations in anti-entropic worlds. The second recognizes the essential role that contingent structure of the actual world plays in the development and application of our causal concepts and reasoning strategies and, in doing so, touches on some of the central ideas of \citep{2020_woodward}.

I have employed a difference-making notion of causation throughout. One could abandon such a notion in favor of one according to which, for example, causation is a matter of nomological determination \citep{2009_ney} or the transfer of conserved quantities \citep{1984_salmon,1994_salmon}, \citep{2000_dowe}. These approaches to causation face severe difficulties \citep[chapter 1]{1998_hausmann}, \citep{2002_hausmann}, \citep{2013_glynn}, \citep{2013_paul_hall} and the fact that they would allow one to understand time reversal as entailing causation reversal does not seem to me to nearly outweigh those difficulties.\footnote{It is notable that such accounts of causation fail to satisfy any of the quite different sets of criteria for evaluating accounts of causation proposed by \citep[chapter 1]{1998_hausmann}, \citep[chapter 2]{2013_paul_hall}, and \citep{2014_woodward}, respectively.} In fact, I think one philosophical consequence of the methods for inferring causal direction discussed in \citep{2020_woodward} is to pose yet another difficulty for such approaches. By employing statistical independence conditions like CSI and VRI, one can ground a causal asymmetry between states of classical or quantum systems related by time-symmetric dynamical laws \citep{2016_janzing_chaves_scholkopf,2022_williams}. As illustrated above, there is nothing necessarily emergent or non-fundamental about these statistical asymmetries. In classical statistical mechanics, they arise between \textit{microstates} whose evolution is described by a Hamiltonian $\mathcal{H}$; if this does not satisfy the requirement that they occur in a theory ``intended to describe the mechanisms of microphysical interactions'' \citep[p. 749]{2009_ney}, then nothing does. By reducing causation to a relation of nomological determination, such accounts indulge in what \citep[section 11]{2020_woodward} calls the ``cause-in-laws'' picture. As a result, such accounts are insensitive to the causal information encoded in the statistical asymmetries that exist between states related by fundamental dynamical laws. Such accounts face the burden of explaining why one can reliably infer causal direction on the basis of such statistical asymmetries if causation really is just nomological determination, i.e. if all of the causally relevant information is contained in the laws alone.

Alternatively, one could retain a difference-making account of causation and argue that the causal ordering of an entropic world remains fixed under time reversal. One might go about this strategy in a couple different ways. If one adopts a standard understanding of the time reversal operation, then a sequence of states $S(t_1), \, \ldots, \, S(t_n)$ related by the dynamics $\mathcal{D}_{t_n}$ and the sequence of time-reversed states $\mathcal{R}S(- t_n), \, \ldots, \, \mathcal{R}S(- t_1)$ also related by $\mathcal{D}_{-t_n}$ describe, in general, distinct physical processes. Suppose that one takes the direction of causation to be the causal ordering that satisfies independence conditions like CSI and VRI; this is the same in an entropic world and its time-reversed, anti-entropic partner, so the direction of causation is fixed under time reversal. This entails that the direction of time and the direction of causation systematically come apart in anti-entropic worlds: effects will almost always precede their causes. Such an account retains meaningful causal relations in anti-entropic worlds, but at the expense of severing any connection between causal and temporal ordering. This is a huge conceptual cost: that connection is central to the everyday and scientific notion(s) of causation whose explication is, at least in part, the aim of providing a philosophical account of causation in the first place.

An alternative version of this strategy is pursued in \citep{2020_farr}. Farr's particular proposal aims to avoid this conceptual cost by adopting a non-standard treatment of time reversal. The reason Farr adopts this non-standard treatment is that it allows him to identify a world and its time-reverse as alternate descriptions of one and the same world; on this non-standard account, the state $S(t)$ and its time-reverse $\mathcal{R}S(-t)$ are simply alternate descriptions of a single physical state of affairs. That means that $S(t_1), \, \ldots, \, S(t_n)$ and $\mathcal{R}S(- t_n), \, \ldots, \, \mathcal{R}S(- t_1)$ are simply two alternative descriptions of a \textit{single} sequence of instanteous physical states of affairs. The correct causal ordering of that single sequence of physical states of affairs is determined by statistical independence conditions like CSI and VRI. This causal ordering picks out one of the two alternative descriptions $S(t_1), \, \ldots$ or $\mathcal{R}S(- t_n), \, \ldots$ as the correct one, and thereby determines the direction of time.

The problem with this alternative version of the strategy hinges on its treatment of time reversal. Many physical quantities change under time reversal: the spin of an electron, the direction of a magnetic field, the momentum of a physical system, and so on. (As described by many of the references in footnote \ref{trdef}, there is good physical reason to require this.) To avoid the conclusion that a sequence of states and its time-reverse describe distinct physical processes, Farr needs to deny physical significance to properties that are not invariant under time reversal; otherwise an instantaneous state and its time-reverse could not simply be alternate descriptions of a single physical state of affairs. This requires Farr to claim that properties that change under time reversal, like momentum, the direction of a magnetic field, the spin of a quantum system, etc. ``are either (i) not causal, or (ii) not genuine properties of instantaneous states'' \citep[p. 201]{2020_farr}.\footnote{See also: ``[time reversed]-twins differ only in terms of notation: they represent a single possible world, and hence notation that varies under time reversal (such as the direction of velocities) should not be taken to represent a property of the target system'' \citep[p. 191]{2020_farr} and ``any quantities that differ between [time reversed]-twins (such as instantaneous velocity, spin, and so on, as discussed above) can be considered descriptive artefacts that equally correspond to a single time-direction-independent$\ldots$ state of affairs'' \citep[fn. 18]{2020_farr}.} A full examination of the challenges faced by such a proposal is largely orthogonal to my main focus here, but for present purposes I will say the following. Abandoning the standard interpretation of properties like spin and magnetic fields as genuine, causally efficacious physical properties is (i) \textit{ad hoc} -- the only motivation is that it is required to interpet $S(t)$ and $\mathcal{R}S(-t)$ as describing a single physical state of affairs, which is itself only motivated by Farr's goal of maintaining a fixed causal direction under time reversal -- and (ii) seems to have enormous (but unaddressed in \citep{2020_farr}) consequences for the explanatory resources of our physical theories, apparently invalidating the myriad explanations that invoke these ``causally inefficacious’’ quantities.\footnote{Whether quantities that include time derivatives in their definition, like momentum, are properties of instantaneous states is a more subtle question; see \citep{2000_albert,2000_arntzenius,2003_smith}).} 

I think the prospects for retaining meaningful causal claims in anti-entropic worlds are dim. That said, one might start to wonder what is really at stake here. Such worlds are fantastically foreign to us: eggs unscramble; photons are emitted from our retinas with wavelengths perfectly matched to their incident surfaces; food is reconstituted in stomachs, regurgitated, and spontaneously recombines; previously dead organic matter is reanimated and undergoes anti-aging; and so on. Life in such worlds defies imagination.\footnote{Indeed, I find the argument in \citep[chapter 4.3]{2007_maudlin} that beings in such worlds would not have conscious experiences at all fairly compelling.} So there are no causal relations in such worlds: who cares?

This brings me to the second avenue of response. Down this second avenue lies an epistemically modest attitude: anti-entropic worlds are so different from the actual world that we have no epistemic warrant for judging the presence \textit{or} absence of causal relations in those worlds based on attempts to apply causal concepts developed to reason and navigate in the actual world. 

In particular, the wordly infrastructure that supports causal reasoning in the actual world -- for example, the widespread satisfaction of independence conditions like CSI and VRI -- is entirely absent in anti-entropic worlds.\footnote{This raises an interesting question about the precise connection between the satisfaction of CSI and VRI and the fact that the actual world is entropic.} This presents us with a dramatic mismatch between the worldly infrastructure of anti-entropic worlds and contexts in which our real-world causal concepts and reasoning strategies can be reasonably applied. This is because the concepts and strategies that we have developed for causal reasoning in the actual world are built to exploit precisely the worldly infrastructure that is absent in anti-entropic worlds. Our causal concepts and strategies are \textit{designed} to be applied to worlds that share certain structural features with ours; in particular, worlds in which CSI and VRI are generically satisfied. Here we make contact with \cite{2020_woodward} whose point, in part, is to remind us that our causal concepts have been developed by human beings -- over a long evolutionary history -- for the purpose of exploiting certain pervasive structural features of the world so they can achieve certain pragmatic and intellectual goals.\footnote{See also \citep{2014_woodward} for how this fact should impact the way we approach theorizing about causation.} These structural features of the world include, among other things, various independence conditions like CSI and VRI that allow for local interventions and modular reasoning.\footnote{See also \citep{1999_hausmann_woodward,2002_cartwright,2004_hausmann_woodward}.} \textit{When those structural features are present}, the causal concepts that we have developed to exploit those structural features enable us to make reliable causal judgments. 

However, when confronted with a world in which those structural features are widely or uniformly absent -- for example, worlds in which CSI and VRI fail dramatically -- then epistemic honesty demands we recognize that we are at a loss. We should not expect the causal concepts and strategies that we have developed to navigate the actual world to be applicable: the worldly infrastructure that licenses their use and supports their successful application is not present. If such a world has causal structure at all, our methods are incapable of diagnosing it. Intelligent beings in such a world, presuming they could exist and that their survival (like ours) would depend on successfully exploiting various dependence relationships in that world, would inevitably develop a radically different set of concepts and strategies for doing so. Only hubris could suggest that we can apply our own parochial set of causal concepts to form reliable judgments about the presence or absence of causal relations in such a world.

Some will be inclined to think that by tying an account of causation to contingent features of the actual world so tightly as to restrict its applicability in this way, one is exhibiting something like a lack of philosophical ambition.\footnote{For example, in \citep[section 3.2]{2013_paul_hall} this is sufficient to make one an ``ontological wimp''.} I disagree. Emphasizing the important role that independence conditions like CSI and VRI have played in shaping our causal concepts and supporting successful causal reasoning is, I think, an essential aspect of any naturalistic analysis of causation. Indeed, any such naturalism ought to be built on a foundation that, among other things, includes an acceptance of the contingency of our causal concepts and strategies for formulating successful causal judgments, and of the pragmatic forces driving their developmental history. An approach to analyzing causation that focuses on a detailed accounting of the worldly infrastructure that shapes and supports our causal reasoning strategies promises to yield a richly informative account of causation, but one that can be sensibly applied to a comparatively small volume of the space of possible worlds. Focusing on how contingent structural features of the actual world ground our causal concepts is not a widely adopted understanding of what it means to be engaged in providing an account of the metaphysics of causation, but I think it offers a promising path for those interested in naturalistic analyses of our causal concepts and reasoning strategies.

\section*{Acknowledgments}

For helpful discussions and comments I would like to thank David Albert, Bob Batterman, Naftali Weinberger, and especially Jim Woodward.

\bibliographystyle{plainnat}
\bibliography{causal_asymmetry}

\end{document}